\documentclass[letterpaper,twoside,twocolumn,english,aps,sort,amsfonts,amssymb,amsmath,prb,floatfix,preprintnumbers,showpacs]{revtex4}
\usepackage{amsmath}
\usepackage{graphicx}
\usepackage{amssymb}

\makeatletter

\newcommand{\noun}[1]{\textsc{#1}}
\providecommand{\tabularnewline}{\\}

\usepackage{graphicx}
\usepackage{dcolumn}
\usepackage{bm}

\usepackage{babel}
\makeatother
\begin{document}

\title{The role of carbon for superconductivity in MgC$_{x}$Ni$_{3}$ from
specific heat}

\author{A. W\"alte}

\email{waelte@ifw-dresden.de}

\author{G. Fuchs}

\author{K.-H. M\"uller}

\author{S.-L. Drechsler}

\author{K. Nenkov}

\author{L. Schultz}

\affiliation{Leibniz-Institut f\"ur Festk\"orper- und Werkstoffforschung Dresden,
Postfach 270116, D-01171 Dresden, Germany}

\date{\today}


\begin{abstract}
The influence of carbon deficiency on superconductivity of $\textrm{MgCNi}_{3}$
is investigated by specific heat measurements in the normal and superconducting
state. In order to perform a detailed analysis of the normal state
specific heat, a computer code is developed which allows for an instantaneous
estimate of the main features of the lattice dynamics. By analyzing
the evolution of the lattice vibrations within the series and simultaneously
considering the visible mass enhancement, the loss in the electron-phonon
coupling can be attributed to significant changes of the prominent
Ni vibrations. The present data well supports the recently established
picture of strong electron-phonon coupling and ferromagnetic spin
fluctuations in this compound.
\end{abstract}
\maketitle
The discovery of superconductivity in $\textrm{MgCNi}_{3}$ has caused
much attention,\cite{he01} since the large Ni content suggests a
magnetic state rather than superconductivity. Indeed, up to now a
lot of experimental and theoretical publications point to a ferromagnetic
instability at temperatures near and below the superconducting transition
temperature $T_{\textrm{c}}\approx7\textrm{ K}$.\cite{rosner02,singh01,singer01,shan03,waelte04}
So far there is much discussion on the carbon content in $\textrm{MgCNi}_{3}$,
since only samples with carbon excess of $\approx50$ \% show large
$T_{\textrm{c}}$ values, whereas for the stoichiometric composition
a strongly reduced $T_{\textrm{c}}$ is found,\cite{amos02,ren02}
which may be triggered by enhanced pair-breaking due to increasing
spin fluctuations.\cite{joseph05} However specific heat measurements
indicate that this is not the case but instead the electron-phonon
coupling is significantly reduced,\cite{shan03} accompanied by a
considerable hardening of low-energy vibrations.\cite{johannes04}
Band structure calculations predict a constant\cite{joseph05} or
decreasing\cite{shan03} electron density of states at the Fermi level
(EDOS), leaving the interesting prospect of significantly changing
lattice dynamics. Recent measurements of the carbon isotope effect
support this picture.\cite{klimczuk04a}

\begin{figure}
\vspace{2mm}
\begin{center}\includegraphics[%
  width=0.47\textwidth,
  keepaspectratio]{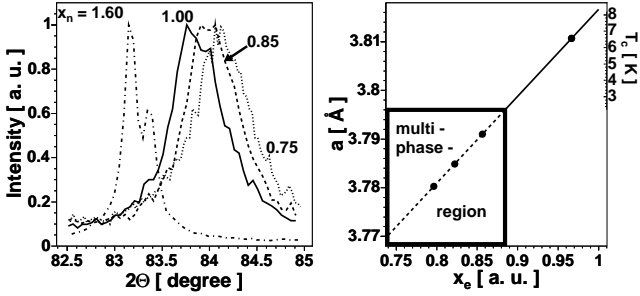}\end{center}

\caption{Structural influence of nominal carbon content on $\textrm{MgC}_{x}\textrm{Ni}_{3}$.
Left panel: x-ray diffractogram around $2\Theta\approx84\textrm{ deg}$.
Right panel: relation between lattice constant $a$, effective carbon
content $x_{\textrm{e}}$ and $T_{\textrm{c}}$ according to Ref.
\onlinecite{amos02} for the investigated samples.\label{fig:misch-xray}}
\end{figure}

Polycrystalline samples of $\textrm{MgCNi}_{3}$ have been prepared
by solid-state reaction. The high volatility of $\textrm{Mg}$ was
balanced in the usual way by $\textrm{Mg}$ excess. Three samples
of nominal composition $\textrm{Mg}_{1.2}\textrm{C}_{x_{\textrm{n}}}\textrm{Ni}_{3}$
with $x_{\textrm{n}}=0.75$, $0.85$ and $1.00$ ($\textrm{n}$=nominal)
have been prepared. For comparison the previously reported sample
with $x_{\textrm{n}}=1.60$ and $T_{\textrm{c}}=6.8\textrm{ K}$ is
chosen.\cite{waelte04} Details on the preparation procedure have
been published elsewhere.\cite{he01} The obtained samples were characterized
by x-ray diffractometry. Due to the relatively low carbon content
no additional graphite was found in the samples, instead a small fraction
of $\textrm{MgNi}_{2}$ forms with decreasing carbon content. However
the fraction never exceeds $6$ Vol.-\% and has minor influence on
the following specific heat analysis. The positions of the reflexes
in the x-ray diffractograms are shifted to higher angles for decreasing
carbon content. The left panel of Fig. \ref{fig:misch-xray} shows
a magnification of the region $2\Theta\approx84\textrm{ deg}$, where
the influence of the carbon deficiency is visible. For a full diffractogram
of the $x_{\textrm{n}}=1.60$ sample see Ref. \onlinecite{waelte04}.
The lattice constants of the samples were determined using the Rietveld
code \noun{fullprof}.\cite{rodriguez90} From that the effective
carbon content was estimated according to Ref. \onlinecite{amos02}.
The dependence is plotted in the right panel of Fig. \ref{fig:misch-xray}
together with the estimated superconducting transition temperatures.\cite{amos02}
It is obvious that the present samples with $x_{\textrm{n}}\leq1.00$
are multi-phase samples. However this is of minor importance for the
following analysis. The specific heat was measured for $T=2-300\textrm{ K}$
and magnetic fields up to $\mu_{0}H=12\textrm{ T}$ using a Quantum
Design Physical Property Measurement System. For the sample with $x_{\textrm{n}}=1.00$,
which shows a considerably broadened superconducting transition (due
to its multi-phase nature) the measurement was extended down to $T=0.3\textrm{ K}$.

\section{Computational details}

\begin{figure}
\vspace{2mm}
\begin{center}\includegraphics[%
  width=0.47\textwidth,
  keepaspectratio]{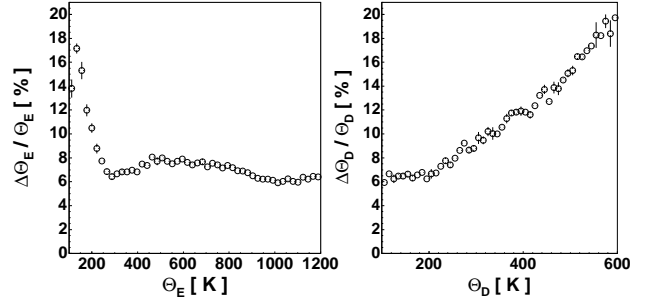}\end{center}

\caption{Energy resolution $\Delta\Theta_{\textrm{D}i}/\Theta_{\textrm{D}i}$
and $\Delta\Theta_{\textrm{E}i}/\Theta_{\textrm{E}i}$ of the code
in a background of $6\left(5\right)$ random Einstein and $1\left(2\right)$
random Debye terms. Left panel: energy resolution for one Einstein
vibration. Right panel: energy resolution for one Debye vibration.\label{fig:energy-resolution}}
\end{figure}

From the mathematical point of view, the extraction of the phonon
density of states (PDOS) from specific heat measurements is ill-posed.
This means, that a number of very different PDOS can lead to very
similar specific heat curves. From the experimental point of view
one has to extract the phonon specific heat from the electron and
other non-phononic background exactly. In reality temperature dependent
effects (e.g. softening) then prevent an inversion approach in most
cases. From this starting point we developed a computer code based
on simple models, namely a Debye and an Einstein model, in order to
get a rough estimate on the lattice dynamics from specific heat measurements
instead of a mathematical exact inversion. Specifications of the code,
which is fully based on \noun{root},\cite{brun97} will be published
elsewhere. Here we want to sketch only the main ideas and illustrate
the capabilities of the code. The specific heat of a compound with
$n$ atoms per unit cell has $3n-3$ optical and $3$ acoustic vibrations
which can be approximated by a linear combination of Einstein and
Debye models. The contribution of the Einstein phonons to the specific
heat is given by:\[
c_{\textrm{E}}\left(T\right)=\sum_{i=3}^{N}R\left(\frac{\Theta_{\textrm{E}i}}{T}\right)^{2}\frac{\exp\left(\Theta_{\textrm{E}i}/T\right)}{\left[\exp\left(\Theta_{\textrm{E}i}/T\right)-1\right]^{2}},\]
with number of optical modes $N$ and Einstein temperatures $\Theta_{\textrm{E}i}$.
The Debye contribution reads:\[
c_{\textrm{D}}\left(T\right)=\sum_{i=0}^{2}3R\left(\frac{T}{\Theta_{\textrm{D}i}}\right)^{3}\int_{0}^{\Theta_{\textrm{D}i}/T}\textrm{d}x\frac{\textrm{e}^{x}x^{4}}{\left(\textrm{e}^{x}-1\right)^{2}},\]
with Debye temperatures $\Theta_{\textrm{D}i}$. %
The form of the phonon density of states is thereby simplified as:\[
F\left(\omega\right)=3R\omega^{2}\sum_{i=0}^{2}\frac{\theta\left(\omega_{\textrm{D}i}-\omega\right)}{\omega_{\textrm{D}i}^{3}}+\sum_{i=3}^{N}\frac{\exp\left[-\frac{\left(\omega-\omega_{\textrm{E}i}\right)^{2}}{2\sigma_{i}^{2}}\right]}{\sigma_{i}\sqrt{2\pi}},\]
with step function $\theta\left(x_{0}-x\right)$ and the characteristic
temperatures in $\textrm{meV}$, indicated by $\omega_{\textrm{D}i}$
and $\omega_{\textrm{E}i}$. For the case of $\textrm{MgCNi}_{3}$
one is left with $3$ Debye and $12$ Einstein terms, adding up to
$15$ parameters. To avoid the need of restrictions on the parameters
we developed a fast converging algorithm. It makes use of a library
containing integrated values of a single Debye and Einstein model
for $\Delta T=1\textrm{ K}$ temperature intervals between $T=2\textrm{ K}$
and $300\textrm{ K}$ and the overall integral of the two models in
the interval $T=2-300\textrm{ K}$. The algorithm integrates the measured
specific heat in a number of temperature intervals and once over the
measurement range $T=2-300\textrm{ K}$. The use of the overall integral
for the fitting procedure accelerates the procedure considerably and
most importantly allows the omission of parameter-restrictions. The
code was statistically tested by randomly distributing $6$ Einstein
and $2$ Debye terms. The fitting result was than compared to the
starting conditions. Fig. \ref{fig:energy-resolution} shows the resulting
energy resolution gained from $20000$ analyzed specific heat model-curves.%
This test shows that the energy resolution is weak for low-energy
optical and high-energy acoustic vibrations, which are both rare cases.
However, there is still potential in the algorithm to improve the
energy resolution. For example by using a set of weighting factors
for different temperature regions, which we are currently working
on.

\section{Analysis and discussion}

\begin{figure}
\vspace{2mm}
\begin{center}\includegraphics[%
  width=0.38\textwidth,
  keepaspectratio]{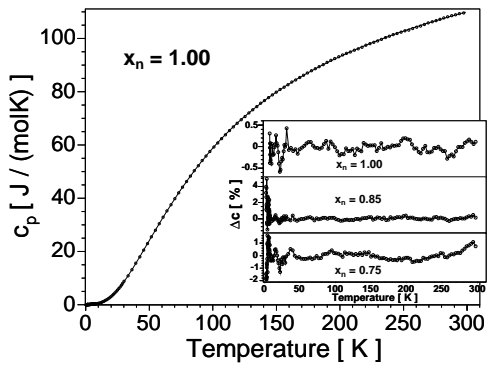}\end{center}

\caption{Specific heat of $\textrm{MgC}_{1.00}\textrm{Ni}_{3}$. Solid line:
specific heat fit (see text for details). Inset: relative difference
between fit and specific heat for $x_{\textrm{n}}=1.00$, $0.85$
and $0.75$ (from top to bottom).\label{fig:cp-300K}}
\end{figure}

The specific heat of the sample with $x_{\textrm{n}}=1.00$ is shown
in Fig. \ref{fig:cp-300K} (the samples with $x_{\textrm{n}}=0.85$
and $0.75$ show very similar results and are omitted for clarity).
The specific heat of $\textrm{MgCNi}_{3}$ is given by a lattice part
$c_{\textrm{lattice}}\left(T\right)$, determined by the above mentioned
code and an electron part $\gamma_{\textrm{N}}T$ with the Sommerfeld
parameter:\begin{equation}
\gamma_{\textrm{N}}=\left[1+\lambda_{\textrm{ph}}+\lambda_{\textrm{sf}}\left(0\right)\right]\gamma_{0},\label{eq:mass-enhancement}\end{equation}
with free electron parameter $\gamma_{0}=11\textrm{ mJ}/\left(\textrm{molK}^{2}\right)$.
The mass enhancement due to the electron-paramagnon coupling can be
estimated from\begin{eqnarray}
\lambda_{\textrm{sf}}\left(T\right) & = & \frac{6}{\pi k_{\textrm{B}}T}\int_{0}^{\infty}\textrm{d}\omega\alpha^{2}F_{\textrm{sf}}\left(\omega\right)\left\{ -z-2z^{2}\Im\left[\psi^{\prime}\left(iz\right)\right]\right.\nonumber \\
 &  & \left.-z^{3}\Re\left[\psi^{\prime\prime}\left(iz\right)\right]\right\} ,\label{eq:paramagnon}\end{eqnarray}
where $\psi\left(iz\right)$ is the digamma function and $z=\omega/\left(2\pi k_{\textrm{B}}T\right)$.
The electron-paramagnon spectral density is given by\[
\alpha^{2}F_{\textrm{sf}}\left(\omega\right)=a\omega\theta\left(\omega_{\textrm{sf}}-\omega\right)+\frac{b}{\omega^{3}}\theta\left(\omega-\omega_{\textrm{sf}}\right).\]
The paramagnon energy, which was estimated in the case of $x_{\textrm{n}}=1.60$
as $\omega_{\textrm{sf}}\approx2.15\textrm{ meV}$,\cite{waelte04}
was also used for the present samples. The black line in Fig. \ref{fig:cp-300K}
is the result of the fitting procedure. The differences between the
data and fit results are shown in the inset for $x_{\textrm{n}}=0.75$,
$0.85$ and $1.00$. They are very small, in particular for $x_{\textrm{n}}=0.85$
and $1.00$, indicating the success of the fitting procedure.

\begin{figure}
\vspace{2mm}
\begin{center}\includegraphics[%
  width=0.38\textwidth,
  keepaspectratio]{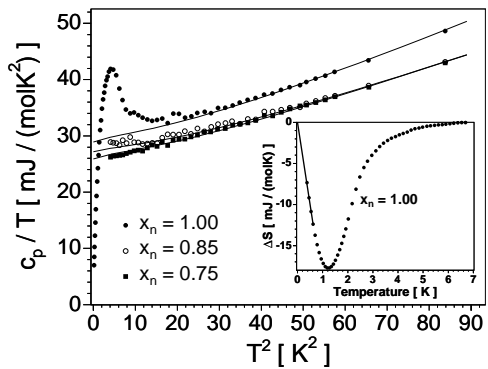}\end{center}

\caption{Specific heat of $\textrm{MgC}_{x}\textrm{Ni}_{3}$ at low temperatures.
Solid lines: fits to the data (see text for details). The anomaly
at $T\approx2.5\textrm{ K}$ for sample $x_{\textrm{n}}=1.00$ indicates
the superconducting transition. Inset: entropy change of the electronic
specific heat within the superconducting state for $x_{\textrm{n}}=1.00$.\label{fig:cp-low-T}}
\end{figure}

Fig. \ref{fig:cp-low-T} shows the low temperature region of the specific
heat. The reduction of $\gamma_{\textrm{N}}$ and the increasing flattening
of the curves with decreasing carbon content was already discovered
in a previous work by \citeauthor{shan03}\cite{shan03} For $x_{\textrm{n}}=1.00$
a strongly broadened superconducting transition is visible at $T_{\textrm{c}}\approx2.5\textrm{ K}$,
in accord with the mentioned multi-phase nature. The deviation from
the low temperature Debye approximation (linear behavior) due to spin
fluctuations is clearly seen for all samples (Fig. \ref{fig:cp-low-T}). 

The model description of the normal state is given as solid lines.
For $x_{\textrm{n}}=1.00$ the conservation of entropy of the superconducting
state, which is expected for a phase transition of second order, was
used as an additional requirement for a successful fit (inset of Fig.
\ref{fig:cp-low-T}). $\gamma_{\textrm{N}}$ was varied with a step-size
of $0.05\textrm{ mJ}/\textrm{molK}^{2}$ until the relative difference
shown in the inset of Fig. \ref{fig:cp-low-T} was minimized. Close
to the final value of $\gamma_{\textrm{N}}$, the derived sets of
vibrational parameters did not appear qualitatively different from
each other. The results of the fitting procedure are summarized in
Tab. \ref{tab:parameters1}. The electron-paramagnon coupling constant
$\lambda_{\textrm{sf}}\left(0\right)$ only slightly decreases by
lowering the carbon content, despite a predicted increase of spin
fluctuations.\cite{joseph05} The electron-phonon coupling constant
shows a strong decrease from $\lambda_{\textrm{ph}}=1.84$ ($x_{\textrm{n}}=1.60$)
down to $\lambda_{\textrm{ph}}=1.06$ ($x_{\textrm{n}}=0.75$), in
agreement with results obtained by \citeauthor{shan03}\cite{shan03}%

\begin{figure}
\vspace{2mm}
\begin{center}\includegraphics[%
  width=0.36\textwidth,
  keepaspectratio]{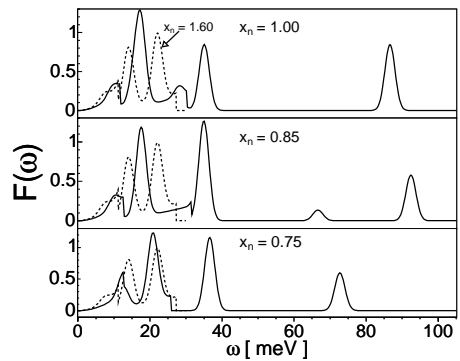}\end{center}

\caption{Phonon density of states of $\textrm{MgC}_{x}\textrm{Ni}_{3}$. Dashed
lines: result for $x_{\textrm{n}}=1.60$.\cite{waelte04}\label{fig:pdos}}
\end{figure}
Fig. \ref{fig:pdos} shows the resulting phonon density of states,
$F\left(\omega\right)$ for $x_{\textrm{n}}=0.75$, $0.85$ and $1.00$
in comparison with previous results for $x_{\textrm{n}}=1.60$.\cite{waelte04}
There are two main changes visible within the series. First there
is a considerable change of the high-energy mode, which is expected,
since it is dominated by carbon. Second all low-energy modes are slightly
shifted to higher energies, accompanied by the shrinking of the unit
cell. The extent of this behavior can be quantified by calculating
the characteristic phonon frequency:\[
\omega_{\textrm{ln}}=\exp\left[\frac{2}{\lambda_{\textrm{ph}}}\int_{0}^{\infty}\textrm{d}\omega\frac{\alpha^{2}\left(\omega\right)F\left(\omega\right)}{\omega}\ln\left(\omega\right)\right],\]
with\begin{equation}
\lambda_{\textrm{ph}}=2\int_{0}^{\infty}\textrm{d}\omega\frac{\alpha^{2}\left(\omega\right)F\left(\omega\right)}{\omega}\label{eq:lambda}\end{equation}
and $\alpha^{2}\left(\omega\right)$ as the electron-phonon interaction
function. In a previous work we suggested to use an approach of the
form $\alpha^{2}\left(\omega\right)=\delta/\sqrt{\omega}$, with scaling
parameter $\delta$ to approximate the electron-phonon interaction
function. Within this approach the low-energy phonons are more strongly
weighted. Calculating $\omega_{\textrm{ln}}$ for the present samples,
an increase from $\omega_{\textrm{ln}}=154\textrm{ K}$ ($x_{\textrm{n}}=1.00$)
to $162\textrm{ K}$ ($x_{\textrm{n}}=0.85$) and $164\textrm{ K}$
($x_{\textrm{n}}=0.75$), compared to $\omega_{\textrm{ln}}=143\textrm{ K}$
for $x_{\textrm{n}}=1.60$ is found.\cite{waelte04} Even by leaving
out the carbon-dominated high-energy mode in the calculation, $\omega_{\textrm{ln}}$
is still increasing {[}$\omega_{\textrm{ln}}=148\textrm{ K}$ ($x_{\textrm{n}}=1.00$),
$\omega_{\textrm{ln}}=154\textrm{ K}$ ($x_{\textrm{n}}=0.85$), $\omega_{\textrm{ln}}=155\textrm{ K}$
($x_{\textrm{n}}=0.75$){]}.%
\begin{table}

\caption{Experimental quantities and electron-boson coupling constants as
derived from the electronic specific heat.\label{tab:parameters1}}

\begin{ruledtabular}\begin{tabular}{lldddd}
&
&
\multicolumn{4}{c}{$\textrm{MgC}_{x}\textrm{Ni}_{3}$}\tabularnewline
\hline
$x_{\textrm{n}}$&
$\left[\textrm{a. u.}\right]$&
1.60&
1.00&
0.85&
0.75\tabularnewline
$x_{\textrm{e}}$&
$\left[\textrm{a. u.}\right]$&
0.967&
0.856&
0.822&
0.796\tabularnewline
$T_{\textrm{c}}^{\textrm{exp}}$&
$\left[\textrm{K}\right]$&
6.80&
\sim2.5\footnotemark[1]&
<2&
<2\tabularnewline
$\gamma_{\textrm{N}}$&
$\left[\textrm{mJ/molK}^{2}\right]$&
31.4&
24.8&
23.5&
22.7\tabularnewline
$\lambda_{\textrm{ph}}$&
&
1.84&
1.25&
1.14&
1.06\tabularnewline
$\lambda_{\textrm{sf}}\left(0\right)$&
&
0.43&
0.38&
0.35&
0.30\tabularnewline
\end{tabular}

\end{ruledtabular}

\footnotetext[1]{with $T_{\rm{c}}^{\rm{onset}}\approx6$ K from ac-susceptibility measurements.}
\end{table}

\begin{figure}
\vspace{2mm}
\begin{center}\includegraphics[%
  width=0.35\textwidth,
  keepaspectratio]{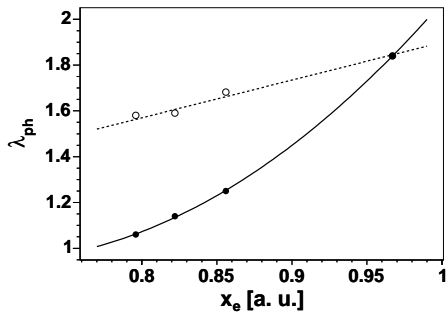}\end{center}

\caption{Electron-phonon coupling constant vs. effective carbon content. Filled
symbols: determined from Eq. (\ref{eq:mass-enhancement}). Open symbols:
calculated from Eq. (\ref{eq:lambda}). The lines are guides to the
eye.\label{fig:epi}}
\end{figure}
This unambiguously demonstrates that the influence of carbon on the
low-energy modes, which are crucial for the observed strong electron-phonon
coupling in $\textrm{MgCNi}_{3}$ (see Ref. \onlinecite{Ignatov03_1})
can not be neglected. A similar result was already derived from structural
investigations.\cite{amos02} In order to stress this point, $\lambda_{\textrm{ph}}$
was calculated from Eq. (\ref{eq:lambda}), using $\delta\approx4.8$
from Ref. \onlinecite{waelte04}. Fig. \ref{fig:epi} shows a comparison
of $\lambda_{\textrm{ph}}$ determined by Eq. (\ref{eq:mass-enhancement})
and Eq. (\ref{eq:lambda}), respectively. From the phonon density
of states a linear dependence from the effective carbon content $x_{\textrm{e}}$
is derived (open symbols), which strongly overestimates the non-linear
result determined from Eq. (\ref{eq:mass-enhancement}) (filled symbols).
This strongly indicates that carbon-deficiency not only influences
the low-energy phonons within the series but also changes the electron-phonon
interaction function $\alpha^{2}\left(\omega\right)$, entering Eq.
(\ref{eq:lambda}).

The present results clearly point to the importance of carbon for
stabilizing the low-energy Ni-dominated phonon modes which are expected
to lead to the high electron-phonon coupling in this compound.\cite{Ignatov03_1}
A similar scenario was established by \citeauthor{johannes04}, who
attributed the phonon hardening in carbon-deficient $\textrm{MgCNi}_{3}$
and the related carbon-deficient $\textrm{ZnCNi}_{3}$ to the low-energy
Ni {}``breathing'' mode.\cite{johannes04} It seems unlikely that
the observed strong decrease of $\lambda_{\textrm{ph}}$ is explained
by the decreasing lattice parameter only, since pressure experiments
show an increase of $T_{\textrm{c}}$ with increasing pressure.\cite{yang03,garbarino04}
Considering a possible decrease of the EDOS,\cite{shan03} one would
expect slightly higher values of $\lambda_{\textrm{ph}}$. However,
considering the well known McMillan formula and the low $T_{\textrm{c}}\lesssim2.5\textrm{ K}$
of the samples with $x_{\textrm{n}}\leq1.00$ this effect seems to
be negligible.\cite{waelte04} The present results may also help to
understand recent investigations of the carbon isotope effect performed
by \citeauthor{klimczuk04a},\cite{klimczuk04a} which is much stronger
than expected from calculations.\cite{Ignatov03_1,dolgov05b} Further
experimental and theoretical studies of the evolution of the EDOS
as well as the PDOS of carbon-deficient and isotope-pure samples and
the behavior of the electron-phonon interaction function $\alpha^{2}\left(\omega\right)$
are highly desirable.

\begin{acknowledgments}
The DFG (SFB 463) is gratefully acknowledged for financial support.
We thank S. V. Shulga for fruitful discussions and the help in fitting
the paramagnon contribution.
\end{acknowledgments}


\begin{thebibliography}{17}
\expandafter\ifx\csname natexlab\endcsname\relax\def\natexlab#1{#1}\fi
\expandafter\ifx\csname bibnamefont\endcsname\relax
  \def\bibnamefont#1{#1}\fi
\expandafter\ifx\csname bibfnamefont\endcsname\relax
  \def\bibfnamefont#1{#1}\fi
\expandafter\ifx\csname citenamefont\endcsname\relax
  \def\citenamefont#1{#1}\fi
\expandafter\ifx\csname url\endcsname\relax
  \def\url#1{\texttt{#1}}\fi
\expandafter\ifx\csname urlprefix\endcsname\relax\def\urlprefix{URL }\fi
\providecommand{\bibinfo}[2]{#2}
\providecommand{\eprint}[2][]{\url{#2}}

\bibitem[{\citenamefont{He et~al.}(2001)\citenamefont{He, Huang, Ramirez, Wang,
  Regan, Rogado, Hayward, Haas, Slusky, Inumara et~al.}}]{he01}
\bibinfo{author}{\bibfnamefont{T.}~\bibnamefont{He}},
  \bibinfo{author}{\bibfnamefont{Q.}~\bibnamefont{Huang}},
  \bibinfo{author}{\bibfnamefont{A.~P.} \bibnamefont{Ramirez}},
  \bibinfo{author}{\bibfnamefont{Y.}~\bibnamefont{Wang}},
  \bibinfo{author}{\bibfnamefont{K.~A.} \bibnamefont{Regan}},
  \bibinfo{author}{\bibfnamefont{N.}~\bibnamefont{Rogado}},
  \bibinfo{author}{\bibfnamefont{M.~A.} \bibnamefont{Hayward}},
  \bibinfo{author}{\bibfnamefont{M.~K.} \bibnamefont{Haas}},
  \bibinfo{author}{\bibfnamefont{J.~S.} \bibnamefont{Slusky}},
  \bibinfo{author}{\bibfnamefont{K.}~\bibnamefont{Inumara}},
  \bibinfo{author}{\bibfnamefont{H.~W.}~\bibnamefont{Zandbergen}},
  \bibinfo{author}{\bibfnamefont{N.~P.}~\bibnamefont{Ong}},
  \bibinfo{author}{\bibfnamefont{R.~J.}~\bibnamefont{Cava}},
  \bibinfo{journal}{Nature}
  \textbf{\bibinfo{volume}{411}}, \bibinfo{pages}{54} (\bibinfo{year}{2001}).

\bibitem[{\citenamefont{Rosner et~al.}(2002)\citenamefont{Rosner, Weht,
  Johannes, Pickett, and Tosatti}}]{rosner02}
\bibinfo{author}{\bibfnamefont{H.}~\bibnamefont{Rosner}},
  \bibinfo{author}{\bibfnamefont{R.}~\bibnamefont{Weht}},
  \bibinfo{author}{\bibfnamefont{M.~D.} \bibnamefont{Johannes}},
  \bibinfo{author}{\bibfnamefont{W.~E.} \bibnamefont{Pickett}},
  \bibnamefont{and} \bibinfo{author}{\bibfnamefont{E.}~\bibnamefont{Tosatti}},
  \bibinfo{journal}{Phys. Rev. Lett.} \textbf{\bibinfo{volume}{88}},
  \bibinfo{pages}{027001} (\bibinfo{year}{2002}).

\bibitem[{\citenamefont{Singh and Mazin}(2001)}]{singh01}
\bibinfo{author}{\bibfnamefont{D.~J.} \bibnamefont{Singh}} \bibnamefont{and}
  \bibinfo{author}{\bibfnamefont{I.~I.} \bibnamefont{Mazin}},
  \bibinfo{journal}{Phys. Rev. B} \textbf{\bibinfo{volume}{64}},
  \bibinfo{pages}{140507(R)} (\bibinfo{year}{2001}).

\bibitem[{\citenamefont{Singer et~al.}(2001)\citenamefont{Singer, Imai, He,
  Hayward, and Cava}}]{singer01}
\bibinfo{author}{\bibfnamefont{P.~M.} \bibnamefont{Singer}},
  \bibinfo{author}{\bibfnamefont{T.}~\bibnamefont{Imai}},
  \bibinfo{author}{\bibfnamefont{T.}~\bibnamefont{He}},
  \bibinfo{author}{\bibfnamefont{M.~A.} \bibnamefont{Hayward}},
  \bibnamefont{and} \bibinfo{author}{\bibfnamefont{R.~J.} \bibnamefont{Cava}},
  \bibinfo{journal}{Phys. Rev. Lett.} \textbf{\bibinfo{volume}{87}},
  \bibinfo{pages}{257601} (\bibinfo{year}{2001}).

\bibitem[{\citenamefont{Shan et~al.}(2003)\citenamefont{Shan, Xia, Liu, Wen,
  Ren, Che, and Zhao}}]{shan03}
\bibinfo{author}{\bibfnamefont{L.}~\bibnamefont{Shan}},
  \bibinfo{author}{\bibfnamefont{K.}~\bibnamefont{Xia}},
  \bibinfo{author}{\bibfnamefont{Z.~Y.} \bibnamefont{Liu}},
  \bibinfo{author}{\bibfnamefont{H.~H.} \bibnamefont{Wen}},
  \bibinfo{author}{\bibfnamefont{Z.~A.} \bibnamefont{Ren}},
  \bibinfo{author}{\bibfnamefont{G.~C.} \bibnamefont{Che}}, \bibnamefont{and}
  \bibinfo{author}{\bibfnamefont{Z.~X.} \bibnamefont{Zhao}},
  \bibinfo{journal}{Phys. Rev. B} \textbf{\bibinfo{volume}{68}},
  \bibinfo{pages}{024523} (\bibinfo{year}{2003}).

\bibitem[{\citenamefont{W{\"a}lte et~al.}(2004)\citenamefont{W{\"a}lte, Fuchs,
  M{\"u}ller, Handstein, Nenkov, Narozhnyi, Drechsler, Shulga, Schultz, and
  Rosner}}]{waelte04}
\bibinfo{author}{\bibfnamefont{A.}~\bibnamefont{W{\"a}lte}},
  \bibinfo{author}{\bibfnamefont{G.}~\bibnamefont{Fuchs}},
  \bibinfo{author}{\bibfnamefont{K.-H.} \bibnamefont{M{\"u}ller}},
  \bibinfo{author}{\bibfnamefont{A.}~\bibnamefont{Handstein}},
  \bibinfo{author}{\bibfnamefont{K.}~\bibnamefont{Nenkov}},
  \bibinfo{author}{\bibfnamefont{V.~N.} \bibnamefont{Narozhnyi}},
  \bibinfo{author}{\bibfnamefont{S.-L.} \bibnamefont{Drechsler}},
  \bibinfo{author}{\bibfnamefont{S.~V.} \bibnamefont{Shulga}},
  \bibinfo{author}{\bibfnamefont{L.}~\bibnamefont{Schultz}}, \bibnamefont{and}
  \bibinfo{author}{\bibfnamefont{H.}~\bibnamefont{Rosner}},
  \bibinfo{journal}{Phys. Rev. B} \textbf{\bibinfo{volume}{70}},
  \bibinfo{pages}{174503} (\bibinfo{year}{2004}).

\bibitem[{\citenamefont{Amos et~al.}(2002)\citenamefont{Amos, Huang, Lynn, He,
  and Cava}}]{amos02}
\bibinfo{author}{\bibfnamefont{T.~G.} \bibnamefont{Amos}},
  \bibinfo{author}{\bibfnamefont{Q.}~\bibnamefont{Huang}},
  \bibinfo{author}{\bibfnamefont{J.~W.} \bibnamefont{Lynn}},
  \bibinfo{author}{\bibfnamefont{T.}~\bibnamefont{He}}, \bibnamefont{and}
  \bibinfo{author}{\bibfnamefont{R.~J.} \bibnamefont{Cava}},
  \bibinfo{journal}{Sol. State Comm.} \textbf{\bibinfo{volume}{121}},
  \bibinfo{pages}{73} (\bibinfo{year}{2002}).

\bibitem[{\citenamefont{Ren et~al.}(2002)\citenamefont{Ren, Che, Jia, Chen, Ni,
  Liu, and Zhao}}]{ren02}
\bibinfo{author}{\bibfnamefont{Z.~A.} \bibnamefont{Ren}},
  \bibinfo{author}{\bibfnamefont{G.~C.} \bibnamefont{Che}},
  \bibinfo{author}{\bibfnamefont{S.~L.} \bibnamefont{Jia}},
  \bibinfo{author}{\bibfnamefont{H.}~\bibnamefont{Chen}},
  \bibinfo{author}{\bibfnamefont{Y.~M.} \bibnamefont{Ni}},
  \bibinfo{author}{\bibfnamefont{G.~D.} \bibnamefont{Liu}}, \bibnamefont{and}
  \bibinfo{author}{\bibfnamefont{Z.~X.} \bibnamefont{Zhao}},
  \bibinfo{journal}{Physica C} \textbf{\bibinfo{volume}{371}},
  \bibinfo{pages}{1} (\bibinfo{year}{2002}).

\bibitem[{\citenamefont{Joseph and Singh}()}]{joseph05}
\bibinfo{author}{\bibfnamefont{P.~J.~T.} \bibnamefont{Joseph}}
  \bibnamefont{and} \bibinfo{author}{\bibfnamefont{P.~P.} \bibnamefont{Singh}},
  \bibinfo{note}{cond-mat/0504659}.

\bibitem[{\citenamefont{Johannes and Pickett}(2004)}]{johannes04}
\bibinfo{author}{\bibfnamefont{M.~D.} \bibnamefont{Johannes}} \bibnamefont{and}
  \bibinfo{author}{\bibfnamefont{W.~E.} \bibnamefont{Pickett}},
  \bibinfo{journal}{Phys. Rev. B} \textbf{\bibinfo{volume}{70}},
  \bibinfo{pages}{060507(R)} (\bibinfo{year}{2004}).

\bibitem[{\citenamefont{Klimczuk and Cava}(2004)}]{klimczuk04a}
\bibinfo{author}{\bibfnamefont{T.}~\bibnamefont{Klimczuk}} \bibnamefont{and}
  \bibinfo{author}{\bibfnamefont{R.~J.} \bibnamefont{Cava}},
  \bibinfo{journal}{Phys. Rev. B} \textbf{\bibinfo{volume}{70}},
  \bibinfo{pages}{212514} (\bibinfo{year}{2004}).

\bibitem[{\citenamefont{Rodriguez-Carvajal}(1990)}]{rodriguez90}
\bibinfo{author}{\bibfnamefont{J.}~\bibnamefont{Rodriguez-Carvajal}},
  \bibinfo{journal}{Abstracts of the Satellite Meeting on Powder Diffraction of
  the XV Congress of the IUCr} p. \bibinfo{pages}{127} (\bibinfo{year}{1990}),
  \bibinfo{note}{Toulouse, France}.

\bibitem[{\citenamefont{Brun and Rademakers}(1997)}]{brun97}
\bibinfo{author}{\bibfnamefont{R.}~\bibnamefont{Brun}} \bibnamefont{and}
  \bibinfo{author}{\bibfnamefont{F.}~\bibnamefont{Rademakers}},
  \bibinfo{journal}{Nucl. Inst. \& Meth. in Phys. Res. A}
  \textbf{\bibinfo{volume}{389}}, \bibinfo{pages}{81} (\bibinfo{year}{1997}),
  \bibinfo{note}{http://root.cern.ch/}.

\bibitem[{\citenamefont{Ignatov et~al.}(2003)\citenamefont{Ignatov, Savrasov,
  and Tyson}}]{Ignatov03_1}
\bibinfo{author}{\bibfnamefont{A.~Y.} \bibnamefont{Ignatov}},
  \bibinfo{author}{\bibfnamefont{S.~Y.} \bibnamefont{Savrasov}},
  \bibnamefont{and} \bibinfo{author}{\bibfnamefont{T.~A.} \bibnamefont{Tyson}},
  \bibinfo{journal}{Phys. Rev. B} \textbf{\bibinfo{volume}{68}},
  \bibinfo{pages}{220504(R)} (\bibinfo{year}{2003}).

\bibitem[{\citenamefont{Yang et~al.}(2003)\citenamefont{Yang, Mollah, Huang,
  Ho, Huang, Liu, Lin, Zhang, Yu, and Jin}}]{yang03}
\bibinfo{author}{\bibfnamefont{H.~D.} \bibnamefont{Yang}},
  \bibinfo{author}{\bibfnamefont{S.}~\bibnamefont{Mollah}},
  \bibinfo{author}{\bibfnamefont{W.~L.} \bibnamefont{Huang}},
  \bibinfo{author}{\bibfnamefont{P.~L.} \bibnamefont{Ho}},
  \bibinfo{author}{\bibfnamefont{H.~L.} \bibnamefont{Huang}},
  \bibinfo{author}{\bibfnamefont{C.-J.} \bibnamefont{Liu}},
  \bibinfo{author}{\bibfnamefont{J.-Y.} \bibnamefont{Lin}},
  \bibinfo{author}{\bibfnamefont{Y.-L.} \bibnamefont{Zhang}},
  \bibinfo{author}{\bibfnamefont{R.-C.} \bibnamefont{Yu}}, \bibnamefont{and}
  \bibinfo{author}{\bibfnamefont{C.-Q.} \bibnamefont{Jin}},
  \bibinfo{journal}{Phys. Rev. B} \textbf{\bibinfo{volume}{68}},
  \bibinfo{pages}{092507} (\bibinfo{year}{2003}).

\bibitem[{\citenamefont{Garbarino et~al.}(2004)\citenamefont{Garbarino,
  Monteverde, N{\'u}{\~n}ez-Regueiro, Acha, Weht, He, Regan, Rogado, Hayward,
  and Cava}}]{garbarino04}
\bibinfo{author}{\bibfnamefont{G.}~\bibnamefont{Garbarino}},
  \bibinfo{author}{\bibfnamefont{M.}~\bibnamefont{Monteverde}},
  \bibinfo{author}{\bibfnamefont{M.}~\bibnamefont{N{\'u}{\~n}ez-Regueiro}},
  \bibinfo{author}{\bibfnamefont{C.}~\bibnamefont{Acha}},
  \bibinfo{author}{\bibfnamefont{R.}~\bibnamefont{Weht}},
  \bibinfo{author}{\bibfnamefont{T.}~\bibnamefont{He}},
  \bibinfo{author}{\bibfnamefont{K.~A.} \bibnamefont{Regan}},
  \bibinfo{author}{\bibfnamefont{N.}~\bibnamefont{Rogado}},
  \bibinfo{author}{\bibfnamefont{M.}~\bibnamefont{Hayward}}, \bibnamefont{and}
  \bibinfo{author}{\bibfnamefont{R.~J.} \bibnamefont{Cava}},
  \bibinfo{journal}{Physica C} \textbf{\bibinfo{volume}{408-410}},
  \bibinfo{pages}{754} (\bibinfo{year}{2004}).

\bibitem[{\citenamefont{Dolgov et~al.}(2005)\citenamefont{Dolgov, Mazin,
  Golubov, Savrasov, and Maksimov}}]{dolgov05b}
\bibinfo{author}{\bibfnamefont{O.~V.} \bibnamefont{Dolgov}},
  \bibinfo{author}{\bibfnamefont{I.~I.} \bibnamefont{Mazin}},
  \bibinfo{author}{\bibfnamefont{A.~A.} \bibnamefont{Golubov}},
  \bibinfo{author}{\bibfnamefont{S.~Y.} \bibnamefont{Savrasov}},
  \bibnamefont{and} \bibinfo{author}{\bibfnamefont{E.~G.}
  \bibnamefont{Maksimov}} (\bibinfo{year}{2005}),
  \bibinfo{note}{cond-mat/0506362v1}.

\end{thebibliography}
\end{document}